\documentclass[11pt]{article}
\usepackage{amsmath}
\usepackage{amsfonts}

\usepackage{amssymb,graphicx}

\usepackage{amssymb}

\newcommand{\koniec}{\begin{flushright}  $\Box $ \end{flushright}}
\newcounter{mnotecount}[section]

\renewcommand{\themnotecount}{\thesection.\arabic{mnotecount}}

\newcommand{\mnote}[1]
{\protect{\stepcounter{mnotecount}}$^{\mbox{\footnotesize
$
\bullet$\themnotecount}}$ \marginpar{
\raggedright\tiny\em
$\!\!\!\!\!\!\,\bullet$\themnotecount: #1} }

\def\tD{\tilde{\Delta}}\def\tth{\tilde{\Theta}}
\def\be{\begin{equation}}
\def\ee{\end{equation}}
\def\bea{\begin{eqnarray}}
\def\eea{\end{eqnarray}}
\def\hg{\hat g}
\def\hw{\hat\omega}
\def\hnabla{\hat\nabla}
\def\hJ{\hat J}

\def\ha{\hat\alpha}
\def\hPhi{\hat\Phi}
\def\hb{\hat\beta}
\def\hc{\hat\gamma}
\def\hW{\hat W}
\def\hz{\hat z}
\def\hu{\hat u}

\def\ho{{\hat o}}
\def\hr{\hat\rho}
\def\hpsi{\hat\psi}

\def\hM{\hat{M}}

\begin{document}

\title{One-sided type-D metrics with aligned Einstein-Maxwell}
\author{Paul Tod\footnote{email: tod@maths.ox.ac.uk }\\Mathematical Institute,\\Oxford University}
%

\maketitle

\begin{abstract}
We consider four-dimensional, Riemannian  metrics for which one or other of the self-dual or anti-self-dual Weyl tensors is type-D and which satisfy the Einstein-Maxwell equations with the corresponding Maxwell field aligned with the type-D Weyl spinor, in the sense of sharing the same Principal Null Directions (or PNDs). Such metrics always have a valence-2 Killing spinor, and therefore a Hermitian structure and at least one Killing vector. We rederive the results of Araneda (\cite{ba}), that these metrics can all be given in terms of a solution of the $SU(\infty)$-Toda field equation, and show that, when there is a second Killing vector commuting with the first, the method of Ward can be applied to show that the metrics can also be given in terms of a pair of axisymmetric solutions of the flat three-dimensional Laplacian. Thus in particular the field equations linearise.



Some examples of the constructions are given.

\end{abstract}
\section{Introduction}

This note is a sequel to \cite{t2}. There we considered four-dimensional, Riemannian, Ricci-flat metrics for which one or other of the self-dual or anti-self-dual Weyl tensors is type-D in the Petrov-Pirani-Penrose classification (for which see e.g. \cite{pr}). It was convenient always to suppose that it was the unprimed Weyl spinor which was type D (and not identically zero) to reduce the number of primed spinor indices in equations. Here we weaken the Ricci-flat condition to allow for Einstein-Maxwell metrics for which the unprimed Maxwell spinor is \emph{aligned} with the unprimed Weyl spinor, in the sense that the PNDs of the Maxwell spinor are both PNDs of the Weyl spinor. We continue to insist that the Weyl spinor which is type D is \emph{not} identically zero.

With the added assumption that the type-D Weyl spinor is \emph{not} identically zero, we'll call these metrics \emph{one-sided type-D with aligned Einstein-Maxwell}. The Lorentzian Kerr-Newman metric has this property and that is where the term \emph{aligned} first arose. Much as our motivation in \cite{t2} came indirectly from the Chen-Teo metric \cite{ct}, here we are thinking about seeking a charged counterpart of Chen-Teo, though so far unsuccessfully. The Chen-Teo metric was found by Aksteiner \cite{ak} to be one-sided type-D, and therefore Hermitian, so it is natural to wonder if it has a charged counterpart with the correspoinding properties, much as the Kerr-Newman solution is related to the Kerr solution.

This problem was considered earlier by Araneda \cite{ba} and the first part of this article re-derives his findings: one-sided type-D metrics with aligned Einstein-Maxwell are obtained by choosing a solution of  the $SU(\infty)$-Toda equation (we'll omit the term ``$SU(\infty)$'' henceforth, and just say ``the Toda equation"), together with the choice of a solution of a monopole-like equation derived from that. It's known that the Toda equation linearises if the solution has an extra symmetry, \cite{w1}, and we saw in \cite{t2} that one-sided type D Ricci-flat metrics with an extra symmetry commuting with the first can be expressed in terms of an axisymmetric solution of the flat three-dimensional Laplacian. This suggests that the field equations for one-sided type-D  metrics with aligned Einstein-Maxwell, and a second symmetry commuting with the first, linearise in much the same way, and we shall see here that they do.

\medskip

Our method is to start in Section 2 with the assumption of a 4-dimensional Riemannian metric subject to three inter-related conditions. For a Ricci-flat metric the three conditions can be seen to  be equivalent, with the aid of the Goldberg-Sachs Theorem (see volume 2 of \cite{pr}) and there is a generalised version of this result in \cite{pr} which helps us to a similar result, Proposition 1, with an aligned Einstein-Maxwell solution. There are quite a few papers in the literature generalising the Goldberg-Sachs theorem, to admit some restricted forms of non-zero Ricci tensor and/or to Riemannian etyrics, see e.g. \cite{ag, pp, ps, rs}, and some contain results close to Proposition 1, but it seemed easier to generate the result within the formalism we work with here.

Then we use the two-component spinor formalism to rederive the expressions of \cite{ba} for the metric in terms of a solution $u$ of the Toda field equation (\ref{k13}) and a solution $W$ of an associated monopole equation (\ref{k12}). The metric automatically has a Killing vector which arises from a valence-2 Killing spinor, which in turn is a consequence of the assumptions on the curvature (by a generalisation of \cite{pw}). Then in Section 3 we add the assumption that there is a second Killing vector, commuting with the first, and deduce that, after possible redefinitions of coordinates and $u$ preserving the Toda field equation, the second Killing vector must be a symmetry of $u$ corresponding to an ignorable coordinate which can be taken to be $y$ in (\ref{k13}). Then in Section 4 we exploit the observation in \cite{w1} that solutions of the Toda field equation independent of $y$ correspond to axisymmetric solutions of the flat 3-dimensional Laplacian. We arrive at our main result: that Riemannian, one-sided type-D 4-metrics with aligned Einstein-Maxwell and two commuting Killing vectors are in one-to-one correspondence with pairs of axisymmetric solutions of the flat three-dimensional Laplacian. The field equations, known to be completely integrable in this case, in fact linearise. 

\medskip


{\bf{Acknowledgements:}} I am grateful to Dr Bernardo Araneda and Prof Maciej Dunajski for useful discussions, and to the Isaac Newton Institute for Mathematical Sciences, Cambridge, for support and hospitality during the programme `Twistors in Geometry and Physics' (TWT) where work on this paper was undertaken. This work was supported by EPSRC grant no EP/R014604/34;

\section{One-sided type-D metrics with aligned Einstein-Maxwell}
In this section, we rederive the results of \cite{ba} in the style of \cite{t1,t2}.  For background on the 2-component spinor formalism see \cite{ht} or \cite {pr}.
\subsection{Some general theory}

Start then with a Riemannian metric $g$ on a 4-manifold $M$. 

\medskip

In \cite{t2} we considered three conditions on such an $(M,g)$ that was already assumed to be Ricci-flat:
\begin{enumerate}
\item $(M,g)$ admits an integrable complex structure compatible with the metric and of the form
\[J_a^{\;b}=J_A^{\;B}\delta_{A'}^{\;B'}\mbox{   with  }J_{AB}:=J_A^{\;C}\epsilon_{CB}=2io_{(A}o^\dagger_{B)},\]
for some spinor field $o_A$ with the normalisation $o_Ao^{\dagger A}=1$.
\item $(M,g)$ admits a valence-2 Killing spinor, which is to say a spinor field $\omega_{AB}=2i\omega o_{(A}o^\dagger_{B)}$ for a real function $\omega$ and a normalised spinor field $o_A$, satisfying
\be\label{ks2}\nabla_{A'(A}\omega_{BC)}=0.\ee
\item the SD Weyl spinor of $(M,g)$ takes the form
\[\psi_{ABCD}=\psi o_{(A}o_Bo^\dagger_Co^\dagger_{D)},\]
for a real function $\psi$ and a normalised spinor field $o_A$ (and the Ricci spinor is zero by assumption).

\end{enumerate}
Then these conditions are all equivalent: condition (1) forces $o^A$ to be geodesic and shear-free (recall this condition is
\be\label{gsf}o^Ao^B\nabla_{AA'}o_B=0,\ee
and henceforth write this condition as \emph{gsf}), and therefore $o^{\dagger A}$ is gsf as well; then the Goldberg-Sachs Theorem \cite{pr} forces condition (3) to hold -- given a 4-dimensional Ricci-flat $M$, a gsf spinor must be a repeated PND of the Weyl spinor, and in the Riemannian setting can only be twice repeated, whence condition (3); and the calculation in \cite{pw} then forces condition (2) from condition (3) for suitable $\omega$ (in fact $\omega$ is a constant multiple of $\psi^{-1/3}$); finally condition (2) forces $o^A$ to be gsf which implies integrability of $J_a^{\;b}$ in (1).

\medskip

In this sequel to \cite{t2}, we wish to modify the assumptions to allow a non-zero Ricci spinor of a particular form. We replace the assumption of Ricci-flatness by the Einstein-Maxwell condition, that there are two Maxwell fields of opposite duality
\[F^{(-)}_{ab}=\phi_{AB}\epsilon_{A'B'},\mbox{   and   }F^{(+)}_{ab}=\rho_{A'B'}\epsilon_{A'B'},\]
both satisfying the source-free Maxwell equations:
\be\label{em0}\nabla^{AA'}\phi_{AB}=0=\nabla^{AA'}\rho_{A'B'},\ee
and the Ricci spinor is expressed in terms of them via
\be\label{em1}\Phi_{ABA'B'}=\phi_{AB}\rho_{A'B'}.\ee
It necessarily follows that the Ricci scalar is constant, and we shall set it to zero.

\medskip

Now we consider three modified conditions and show that they are again equivalent.

\medskip

\noindent{\bf{Proposition 1}}

\medskip

\noindent Consider the following three conditions on a 4-dimensional Riemannian Einstein-Maxwell space $(M,g)$:

\medskip

1$'$. $(M,g)$ admits an integrable complex structure compatible with the metric and of the form
\[J_a^{\;b}=J_A^{\;B}\delta_{A'}^{\;B'}\mbox{   with  }J_{AB}:=J_A^{\;C}\epsilon_{CB}=2io_{(A}o^\dagger_{B)},\]
for some spinor field $o_A$ with the normalisation $o_Ao^{\dagger A}=1$.\emph{ Furthermore}, the Ricci tensor is $J$-invariant (this is what enforces `aligned').

2. $(M,g)$ admits a valence-2 unprimed Killing spinor, so again this is a spinor field $\omega_{AB}=2i\omega o_{(A}o^\dagger_{B)}$ for a real function $\omega$ and a normalised spinor field $o_A$, satisfying
\be\label{ks2}\nabla_{A'(A}\omega_{BC)}=0.\ee

3$'$.  the SD Weyl spinor of $(M,g)$ takes the form
\[\psi_{ABCD}=\psi o_{(A}o_Bo^\dagger_Co^\dagger_{D)},\]
for a real function $\psi$ and a normalised spinor field $o_A$, and the Ricci spinor takes the form
\[\Phi_{ABA'B'}=\phi_{AB}\rho_{A'B'}\]
where $\phi_{AB}$ and $\rho_{A'B'}$ both satisfy Maxwell's equations:
\[\nabla^{AA'}\phi_{AB}=0=\nabla^{AA'}\rho_{A'B'},\]
and $\phi_{AB}$ is aligned with the Weyl spinor in the sense that
\[\phi_{AB}=2i\phi o_{(A}o^\dagger_{B)},\]
for some real function $\phi$.

\medskip

\noindent{\bf{Proof}}: 

\noindent{\bf{(1$'$) implies (3$'$)}}:  Starting from (1$'$), this still implies that $o_A$ is gsf, but we need a generalisation of the Goldberg-Sachs Theorem to bring in condition (3$'$). This is available in Proposition 7.3.35 of volume 2 of \cite{pr}\footnote{As noted in the Introduction, there have been several papers generalising the Goldberg-Sachs theorem, to a Riemannian setting or to allow varieties of nonzero Ricci tensor, e.g. \cite{ag, ps,rs}, and results similar to this Proposition can be found in this literature.}. Adapted to the Riemannian case this asserts:

\medskip

{\it{Prop 7.3.35: Of the following three conditions:

\medskip

(i) $o^A$ is a twice repeated PND of $\psi_{ABCD}$;

(ii) $o^A$ is gsf.

(iii) $o^Ao^Bo^C\nabla^{DA'}\psi_{ABCD}=0.$

\medskip

(i) and (ii) together imply (iii); (i) and (iii) together imply (ii); (ii) and (iii) together imply (i)}}.

\medskip

This will give us what we want. In (1$'$)  the assumption that the Ricci tensor is $J$-invariant implies that $\phi_{AB}$ in (\ref{em1}) is proportional to $J_{AB}$
\be\label{em2}\phi_{AB}=\phi J_{AB}=2i\phi o_{(A}o^\dagger_{B)},\ee
for some real function $\phi$. Now the Bianchi identity is
\[\nabla^{AA'}\psi_{ABCD}=\nabla_B^{B'}\Phi_{CDA'B'}=\rho_{A'B'}\nabla_B^{B'}\phi_{CD},\]
using (\ref{em1}) and (\ref{em0}). Contract this with $o^Bo^Co^D$ to find that the right-hand-side vanishes by the gsf condition on $o^A$, therefore so does the left-hand-side, which is condition (iii) of Prop 7.3.35; we already have (ii), by virtue of (1$'$), so we may conclude that (i) holds: $o^A$ is a repeated PND of $\psi_{ABCD}$, and we have proved condition (3$'$).

\medskip

\noindent{\bf{(2) follows from (1$'$) and (3$'$)}} and therefore from (1$'$) alone. By assumption, we have the Maxwell equation on $\phi_{AB}$:
\[0=\nabla^{AA'}\phi_{AB}=i\nabla^{AA'}(\phi (o_Ao^\dagger_B+o_Bo^\dagger_A)).\]
Contract this with $-o^B$ to obtain
\be\label{m1}-o^A\nabla_{AA'}\phi+2\phi o^Bo^{\dagger A}\nabla_{AA'}o_B=0.\ee
Now consider the Killing spinor equation:
\[0=\nabla_{A'(A}\omega_{BC)}=2i\nabla_{A'(A}\omega o_Bo^\dagger_{C)}.\]
The contraction of this with $o^Ao^Bo^C$ is satisfied if $o^A$ is gsf (which by (1$'$) it is) so it just remains to impose the contraction with $o^Ao^Bo^{\dagger C}$ (the vanishing of all other components then follows by complex conjugation):
\be\label{ks1}o^Ao^Bo^{\dagger C}\nabla_{A'(A}(2i\omega o_Bo^\dagger_{C)})=\frac{2i}{3}(-o^A\nabla_{AA'}\omega-\omega o^Bo^{\dagger A}\nabla_{AA'}o_B).\ee
Compare this with (\ref{m1}): given $\phi$ satisfying (\ref{m1}), we may take $\omega=\phi^{-1/2}$ to satisfy (\ref{ks1}), and we have a Killing spinor.

\medskip

\noindent{\bf{(1$'$) and (3$'$) both follow from (2)}}: This needs some more theory. We'll  assume (2) and Einstein-Maxwell, with the scalar curvature zero. From (\ref{ks2}) we have
\be\label{ks3}
\nabla_{AA'}\omega_{BC}=\epsilon_{AB}K_{CA'}+\epsilon_{AC}K_{BA'},\ee
for a vector $K^a$. Differentiate again and symmetrise
\[\nabla_{A'(A}\nabla^{A'}_{B)}\omega_{CD}=2\psi_{EAB(C}\omega^E_{\;D)}\]
but also 
\[=2\nabla^{A'}_{(A}\epsilon_{B)(C}K_{D)A'}.\]
Deduce that
\[\psi_{E(ABC}\omega^E_{\;D)}=0,\]
which is part of (3$'$) (type-D-ness of the Weyl spinor), and that
\be\label{k1}\nabla_{A'(A}K^{A'}_{B)}=-\frac12\psi_{ABCD}\omega^{CD},\ee
which we use below, and
\[\nabla_aK^a=0,\]
which is the trace-part of the Killing equation on $K^a$.

Bringing in the Ricci spinor we have
\[\nabla_{A(A'}\nabla^A_{\;B')}\omega_{CD}=2\Phi_{A'B'E(C}\omega^E_{\;D)}\]
but also
\[=-2\nabla_{(D(A'}K_{B')C)}.\]
Condition (3$'$) on the Ricci spinor requires this to vanish but we don't have that yet. When we do then this is the trace-free part of the Killing equation on $K^a$ and we may deduce that $K^a$ is a Killing vector.

Note from condition (2) that
\[\omega_{AB}\omega^{AB}=2\omega^2\]
and then contract (\ref{ks3}) with $\omega^{BC}$ to find
\[2\omega\nabla_{AA'}\omega=\omega^{BC}\nabla_{AA'}\omega_{BC}=-2\omega_A^{\;C}K_{CA'}=-2\omega J_A^{\;C}K_{CA'},\]
i.e.
\be\label{k2} \nabla_{AA'}\omega=-J_A^{\;C}K_{CA'}\mbox{   or   }      \nabla_a\omega=K^cJ_{ca},\ee
so that the function $\omega$ is the Hamiltonian for the Killing vector $K^a$, and of course $K^a\nabla_a\omega=0$.

We can complete the proof that condition (2) implies condition (3$'$) by exploiting the familiar fact that the presence of a Killing spinor implies that the metric $g$ is conformal to K\"ahler (\cite{dt,P}): one rescales with the conformal factor $\Omega=\omega^{-1}$:
\be\label{c0}\hg_{ab}=\Omega^2g_{ab}=\omega^{-2}g_{ab},\ee
and rescales the Killing spinor according to
\[\hw_{AB}=\Omega^2\omega_{AB}\]
as this scaling preserves the Killing spinor equation, and the Killing spinor becomes the K\"ahler form for $(\hM,\hg)$. The normalised PNDs of the Killing spinor rescale according to
\[\ho_A= \Omega^{1/2}o_A,\;\ho^A=\Omega^{-1/2}o^A.  \]

Since the K\"ahler form is parallel we have
\[ 0=\hnabla_{A'(A}\hnabla^{A'}_{B)}\hw_{CD}=\hpsi_{CDE(A}\hw^E_{\;D)}  \]
and
\[ 0=\hnabla_{A(A'}\hnabla^{A}_{B')}\hw_{CD}=\hPhi_{A'B'E(C}\hw^E_{\;D)}.  \]
The first of these forces $\hpsi_{ABCD}$ to be type D and since the Weyl spinor is unchanged under conformal rescaling the same is then true of $\psi_{ABCD}$ - this is the part of condition (3$'$) that we already have. The second forces the hatted Ricci spinor to take the form
\[\hPhi_{ABA'B'}=\hw_{AB}\hr_{A'B'},\]
a form that we want for the unhatted Ricci spinor to complete condition (3$'$) (and the $J$-invariance of the Ricci tensor which is part of condition (1$'$)). Under conformal rescaling the Ricci spinor changes according to
\be\label{c1}\hPhi_{ABA'B'}=\Phi_{ABA'B'}-\nabla_{A'(A}\Upsilon_{B)B'}+\Upsilon_{A'(A}\Upsilon_{B)B'},\ee
where $\Upsilon_a=\Omega^{-1}\nabla_a\Omega=-\omega^{-1}\nabla_a\omega$. Contract (\ref{c1}) with $\Omega\ho^A\ho^B=o^Ao^B$ then the left-hand-side vanishes so that
\be\label{k5}0=o^Ao^B(\Phi_{ABA'B'}+\omega^{-1}\nabla_{AA'}\nabla_{BB'}\omega).\ee
Now consider the second term in this:
\[o^Ao^B\nabla_{AA'}\nabla_{BB'}\omega=o^A\nabla_{AA'}(o^B\nabla_{BB'}\omega)-(o^A\nabla_{AA'}o^B)\nabla_{BB'}\omega,\]
\[=o^A\nabla_{AA'}(-o^BJ_B^{\;C}K_{CB'})+(o^A\nabla_{AA'}o^B)J_B^{\;C}K_{CB'}   \]
using (\ref{k2}). The gsf condition (\ref{gsf}) implies
\[o^A\nabla_{AA'}o_B=o_B\alpha_{A'}\]
for some $\alpha_{A'}$, and the definition of $J_A^{\;B}$ implies
\[o^AJ_A^{\;B}=-io^B,\]
so that
\[o^Ao^B\nabla_{AA'}\nabla_{BB'}\omega=io^A\nabla_{AA'}(o^CK_{CA'})-i\alpha_{A'}o^CK_{CB'}=io^Ao^C\nabla_{AA'}K_{CC'}\]
which is zero by (\ref{k1}). From (\ref{k5}) this means that
\[o^Ao^B\Phi_{ABA'B'}=0,\]
and so
\[\Phi_{ABA'B'}=2i\phi o_{(A}o^\dagger_{B)}\rho_{A'B'}\]
for a suitable $\phi$ (in fact $\phi=\omega^{-2}$), and we have recovered condition (3$'$) from condition (2) (this $\phi_{AB}$ can be seen to satisfy the Maxwell equations since it is the rescaling of $\hw_{AB}$, which trivially satisfies the Maxwell equations in $\hM$, by the conformal factor $\Omega^{-1}$ which is the correct rescaling to preserve the Maxwell equations). We also have that the Ricci spinor is $J$-invariant, which completes the proof of condition (1).
\koniec

That completes the proof the Proposition. We turn to the study of the metric and curvature subject to the three conditions in the Proposition.
\subsection{The metric and curvature for $M$}

We make the metric ansatz in the usual way. We introduce a coordinate $t$ so that $K^a\partial_a=\partial_t$ and then $K_adx^a= W^{-1}(dt+A)$ for a one-form $A$ and scalar $W=(g_{ab}K^aK^b)^{-1}$. We introduce mutually orthogonal unit-length one-forms
\[\theta^0=W^{1/2}K=W^{-1/2}(dt+A),\;\theta^1=J\theta^0=-W^{1/2}d\omega\]
and then on the two-plane orthogonal to the span of $(\theta^0,\theta^1)$
\[\theta^2+i\theta^3=W^{1/2}e^{u/2}(dx+idy),\]
with a presently-unknown function $u(x,y,z)$.
It's convenient to set $z=-\omega$, and the metric is then
\be\label{g1}
g=W(e^u(dx^2+dy^2)+dz^2)+W^{-1}(dt+A)^2.\ee
We shall have to calculate the curvature to obtain equations constraining $u,W$ and $A$ but first there is information to be obtained from the observation that this metric is conformal to K\"ahler. The K\"ahler metric, by (\ref{c0}), is
\be\label{k6}\hg=\frac{W}{z^2}(e^u(dx^2+dy^2)+dz^2)+(z^2W)^{-1}(dt+A)^2,\ee
which we may also write as
\[\hg=\hW(e^{\hu}(dx^2+dy^2)+d\hz^2)+\hW^{-1}(dt+A)^2,\]
so that
\be\label{k9}\hW=z^2W,\;e^{\hu}=z^{-4}e^u,\;d\hz^2=dz^2/z^4,\ee
and we'll choose $\hz=-1/z$. The K\"ahler form $\hJ$  is 
\[ \hJ=(dt+A)\wedge d\hz+\hW e^{\hu}dx\wedge dy. \]
We write
\be\label{k8}dA=\ha dy\wedge d\hz+\hb d\hz\wedge dx+\hc dx\wedge dy\ee
for some $\ha,\hb,\hc$ to be found, and then
\[d\hJ=dA\wedge d\hz+d(\hW e^{\hu})\wedge dx\wedge dy=0,\]
gives $\hc=-(\hW e^{\hu})_{\hz}$.

The type (1,0) forms for the metric (\ref{k6}) are spanned by
\[(dt+A)+i\hW d\hz,\;dx+idy\]
so integrability of the complex structure requires just
\[(dA+id\hW\wedge d\hz)\wedge(dx+idy)=0,\]
whence we obtain all of $dA$ in the form of (\ref{k8}) with
\be\label{k7}\ha=-\hW_x,\;\hb=-\hW_y,\;\hc=-(\hW e^{\hu})_{\hz}.\ee
We can express $dA$ in unhatted variables with the aid of (\ref{k9}), (\ref{k8}) and (\ref{k7}) as
\be\label{k10}dA=-W_xdy\wedge dz-W_ydz\wedge dx-z^2\left(\frac{We^u}{z^2}\right)_zdx\wedge dy.\ee
This needs to be closed, which requires
\be\label{k11} W_{xx}+W_{yy}+\left(z^2\left(\frac{We^u}{z^2}\right)_z\right)_z=0,\ee
and this is equation (2.25) in \cite{ba}. It can be simplified slightly by setting $W=zX$ when it becomes
\be\label{k12} X_{xx}+X_{yy}+(Xe^u)_{zz}=0,\ee
and this can conveniently be called \emph{the monopole equation} for $G$.

Along the way we obtain the unprimed Maxwell field in $M$ as
\be\label{m7}\varphi=\frac{1}{z^2}((dt+A)\wedge dz+We^udx\wedge dy),\ee
and this is easily seen to be closed (and therefore also co-closed).

\medskip
%

\medskip

For the curvature, we'll use Cartan calculus on an orthonormal basis of SD 2-forms, so introduce this basis by
\[\phi^1=\theta^0\wedge\theta^1+\theta^2\wedge\theta^3,\;\phi^2=\theta^0\wedge\theta^2+\theta^3\wedge\theta^1, \; \phi^3=\theta^0\wedge\theta^3+\theta^1\wedge\theta^2, \]
and a corresponding basis of ASD 2-forms by
\[\psi^{\bar{1}}=\theta^0\wedge\theta^1-\theta^2\wedge\theta^3,\;\psi^{\bar{2}}=\theta^0\wedge\theta^2-\theta^3\wedge\theta^1, \; \psi^{\bar{3}}=\theta^0\wedge\theta^3-\theta^1\wedge\theta^2. \]
We introduce the connection one-forms $\alpha_i^{\;j}$ by
\[d\phi^i=-\alpha^i_{\;j}\wedge\phi^j\]
(these indices are raised and lowered by $\delta_{ij},\delta^{ij}$) and solve for $\alpha^i_{\;j}$ to find
\[\alpha^1_{\;2}=C\theta^2,\;\alpha^3_{\;1}=-C\theta^3,\;\alpha^2_{\;3}=E\theta^0+G\theta^2+H\theta^3,\]
with
\[C=-W^{-1/2}z^{-1},\;E=W^{-1/2}(\frac{u_z}{2}-\frac{1}{z}),\;G=\frac12W^{-1/2}e^{-u/2}u_y,\;H=-\frac12W^{-1/2}e^{-u/2}u_x.\]


Now obtain the curvature components from
\[ \Omega^i_{\;j}=d\alpha^i_{\;j}+\alpha^i_{\;k}\wedge\alpha^k_{\;j}=  \Omega^i_{\;j\cdot k}\phi^k+\Omega^i_{\;j\cdot\bar{k}}\psi^{\bar{k}}.\]

With our curvature assumptions, namely zero Ricci scalar, type-D SD Weyl spinor and (\ref{em1}) for the Ricci spinor, we may identify
\[\Omega^1_{\;2.k}=E_{33}\delta^3_k,\;\Omega^2_{\;3.k}=E_{11}\delta^1_k,\;\Omega^3_{\;1.k}=E_{22}\delta^2_k,\]
other independent components zero, where $E_{22}=E_{33}=-\frac12E_{11}=\psi/6$ are the nonzero components of the SD Weyl spinor as an endomorphism on SD 2-forms, and
\[\Omega^1_{\;2.\bar{k}}=\Omega^3_{\;1.\bar{k}}=0,\]
with $\Omega^2_{\;3.\bar{k}}$ equal to the components of $\rho_{A'B'}$ in the basis $\{\psi^{\bar{k}}\}$.
Calculating these, we obtain the $\Omega^2_{\;3.\bar{k}}$, which we have no interest in, but also the Toda field equation
\be\label{k13}u_{xx}+u_{yy}+(e^u)_{zz}=0,\ee
and an expression for $\psi$:
\be\label{k14}
\psi =6E_{22}=\frac{1}{2Wz^2}(2-zu_z).\ee

\medskip

To summarise, with the metric form (\ref{g1}), we choose a solution $u$ of (\ref{k13}) and then a solution $X$ of (\ref{k12}); set $W=zX$ and solve (\ref{k10}) for $A$ and substitute into (\ref{g1}). The remaining unknown component of the unprimed Weyl spinor is given by (\ref{k14}), and the aligned Maxwell field by (\ref{m7}).
%

%
 \section{A second Killing vector and the Ward transformation}
 If there is a second Killing vector, then by the same argument as in \cite{t2}, there is no loss of generality is assuming it is $L=\partial_y$ and the functions $u,W$ and $X$ and the components of $A$ are independent of $y$. As in \cite{t2}, we can apply the transformation of \cite{w1} to linearise (\ref{k13}) and simplify (\ref{k12}).

Starting from (\ref{k13}) with $u$ independent of $y$:
\be\label{w0}u_{xx}+(e^u)_{zz}=0,\ee
we change from coordinates $(x,z)$ to coordinates $(R,Z)$ and from dependent function $u(x,z)$ to dependent function $V(R,Z)$ via
\be\label{w00}x=V_Z,\;z=\frac12RV_R,\;u=\log(R^2/4)\mbox{   so that   }R^2=4e^u,\ee
with the assumption that $V$ is axisymmetric and harmonic:
\be\label{w1}(RV_R)_R+RV_{ZZ}=0,\ee
(and that its second derivatives $V_{ZZ},V_{RZ}$ are not both zero).
The Jacobian matrix of the coordinate transformation, taking account of (\ref{w1}),  is
\[\frac{\partial(x,z)}{\partial(R,Z)}=\left(\begin{array}{cc}
         V_{ZR}& -\frac12RV_{ZZ}\\
        V_{ZZ}& \frac12RV_{ZR}\\
\end{array}\right),\]
with inverse
\[\frac{\partial(R,Z)}{\partial(x,z)}=\frac{1}{\Delta}\left(\begin{array}{cc}
         \frac12RV_{ZR}& \frac12RV_{ZZ}\\
        -V_{ZZ}& V_{ZR}\\
\end{array}\right),\]
where $\Delta =\frac{1}{2} R((V_{ZR})^2+(V_{ZZ})^2)$.

\medskip

{\bf{Aside:}} I've precisely followed Ward's original presentation of this transformation \cite{w1} but it has a small disadvantage for our purposes: note that, from (\ref{w00}),
\[\Delta dR\wedge dZ=dx\wedge dz,\]
which has the effect of changing the orientation of the spatial metric $h$ in (\ref{w5}) and consequently introducing a sign change in the choice of the canonical Weyl-Papapetrou coordinates later.

\medskip

\noindent Now calculate
\be\label{w2}e^u\frac{\partial u}{\partial z}=-\frac{\partial Z}{\partial x},\;\frac{\partial u}{\partial x}=\frac{\partial Z}{\partial z},\ee
from which (\ref{w0}) follows. For the converse, given $u$ satisfying (\ref{w0}), obtain $Z$ from (\ref{w2}) and take $R=2e^{u/2}$, discover that
\[\frac12R\frac{\partial x}{\partial R}-\frac{\partial z}{\partial Z}=0=R\frac{\partial x}{\partial Z}+2\frac{\partial z}{\partial R},\]
so introduce $V$ with
\[x=V_Z,\;z=\frac12 RV_R,\]
by the first of these and find it is harmonic by the second.

We also want to transform (\ref{k11}) or (\ref{k12}). For this, first note that
\be\label{w5}
h:=dz^2+e^u(dx^2+dy^2)=\frac12R\Delta(dR^2+dZ^2)+\frac14 R^2dy^2,\ee
so for any $\phi(x,z)$ calculate
\[\Delta_h\phi=e^{-u}(\phi_{xx}+(e^u\phi_z)_z)=\frac{2}{R^2\Delta}((R\phi_R)_R+R\phi_{ZZ}),\]
where $\Delta_h$ is the Laplacian for the metric $h$.

We can solve (\ref{k12}) by setting $X=F_z$ where
\[F_{xx}+F_{yy}+(e^uF_z)_z=0,\]
and for an $F$ independent of $y$ this is the Laplace equation for $h$, so $F$ is equivalently found as a harmonic function of $(R,Z)$.

In the metric, we need $W$ as a function of $(R,Z)$ and this is\footnote{This incidentally shows that, having chosen $V$, we choose $F$ to be an arbitrary second harmonic function, but we must not choose $F=V_Z$ or $W$ is identically zero and the construction collapses.}
\be\label{w3}W=zF_z=\frac{RV_R}{2\Delta}(F_ZV_{ZR}-F_RV_{ZZ}).\ee

To complete the metric (\ref{g1}) we need an expression for $A$. For this we need the function $H$ conjugate to $F$, in the sense that
\[F_x=H_z,\;e^uF_z=-H_x,\]
and then it can be checked (by solving (\ref{k10})) that a candidate for $A$ is
\[A=Bdy\mbox{   with   }B=(zH_z-H).\]
We want to express this in terms of $(R,Z)$ so first note that
\[H_R=-\frac12 RF_Z,\;H_Z=\frac12 RF_R\]
and then
\be\label{w4}A= Bdy=\left(\frac{R^2V_R}{4\Delta}(F_RV_{ZR}+F_ZV_{ZZ})-H\right)dy.\ee 
Now we may assemble the metric (\ref{k6}) by taking the spatial metric from (\ref{w5}), $W$ from (\ref{w3}) and $A$ from (\ref{w4}), remembering that $e^u=R^2/4$. The result can be written in the standard toric form as
\be\label{w6}g=  \Omega^2(dR^2+dZ^2)+       \left(\begin{array}{cc}
 dt&dy\\
 \end{array}\right)\left(\begin{array}{cc}
 W^{-1}&BW^{-1}\\
            BW^{-1} & B^2W^{-1}+WR^2/4\\
 \end{array}\right)\left(\begin{array}{c}
 dt\\
            dy\\
 \end{array}\right), \ee
with
\[\Omega^2=\frac{R^2V_R}{4}(F_Z V_{ZR}-F_R V_{ZZ}),\]
\[W=\frac{RV_R}{2\Delta}(F_Z V_{ZR}-F_R V_{ZZ}),\]
\[B= \frac{R^2V_R}{4\Delta}(F_RV_{ZR}+F_ZV_{ZZ})-H, \]
and still
\[ \Delta =\frac{1}{2} R((V_{ZR})^2+(V_{ZZ})^2).\]
Also the determinant of the Gram matrix (i.e. the matrix of inner products of the Killing vectors) is $R^2/4$ so the Weyl-Papapetrou canonical coordinates are $(R/2,Z/2)$.\footnote{As noted in the aside above, strictly speaking these should be $(R/2, -Z/2)$.}



\section{Some examples}
\begin{itemize}
\item {\bf{Riemannian IWP}}

Before imposing the second symmetry, let us consider the case $u=0$, which can't be the subject of the Ward transformation. Now $W=zX$ and 
\[\Delta_0X:=X_{xx}+X_{yy}+X_{zz}=0,\]
while from (\ref{k10})
\[dA=-zX_xdy\wedge dz-zX_ydz\wedge dx-(zX_z-X)dx\wedge dy,\]
and the metric is 
\[g=zX(dx^2+dy^2+dz^2)+(zX)^{-1}(dt+A)^2.\]
This is a special case of a familiar class of metrics, the Riemannian IWP solutions \cite{dh,Wh,Y}: take two harmonic functions $U_1,U_2$ and solve
\[dA=*(U_1dU_2-U_2dU_1),\]
for the 1-form $A$. Here $*$ is the dual in the flat 3-metric, and the RHS is closed by harmonicity of $U_1,U_2$. Then the metric is
\[g_1=U_1U_2(dx^2+dy^2+dz^2)+(U_1U_2)^{-1}(dt+A)^2,\]
so we choose $U_1=X, U_2=z$ to get the case under consideration. These particular Riemannian IWP metrics are neither regular nor ALF nor ALE.

{\bf{Comment}}: Since the Riemannian IWP metrics depend on a choice of two harmonic functions, as do the type-D aligned Einstein-Maxwell metrics given here, one might worry that all the metrics given here were in this class. Here are two arguments why they aren't: first, the obvious $J$, given by
\[J(dx)=dy,\;J(dt+A)=U_1U_2dz\]
is only integrable if either $U_{1x}=U_{1y}=0$ or the same with $U_2$; second, the IWP metrics include the Majumdar-Papapetrou solutions, which are not always type-D, so not always Hermitian.

\item {\bf{Examples with $e^u$ separable}}

This case includes the spherically symmetric solutions which were discussed in \cite{ba} but there's a little bit to add. Choose the separable solution (i.e. separable as a sum; $e^u$ is separable as a product)
\[u=f(x,y)+g(z)=-2\log\cosh x+\log(z^2+2az+b),\]
with $a,b$ constant (the solution
\[u=f(x,y)+g(z)=-2\log(1+x^2+y^2)+\log(4(z^2+2az+b))\]
gives the same result but the one we've chosen is independent of $y$, so is an example for application of the theory in Section 3.). Then choose $W$ as a function only of $z$, for which the general solution is 
\[W=\frac{z(cz+2n)}{(z^2+2az+b)},\]
where $c,n$ are two more constants, and then calculate
\[A=2n\cos\theta d\phi,\]
having introduced angular coordinates $\theta,\phi$ by
\be\label{e1}y=\phi,\;x=\log\tan(\theta/2),\ee
so that also $\cos\theta=\tanh x$ and $\sin\theta=\mbox{sech } x$.
Relabel the constants according to
\[c=k^2,\;n=kN,\;a=-\frac{1}{k}(m-N),\;b=\frac{1}{k^2}(2N^2-2mN+e^2),\]
and introduce $r,\chi$ by
\[r=kz+N,\;t=k\chi\]
then the metric (\ref{g1}) becomes
\[g=\frac{dr^2}{U(r)}+(r^2-N^2)(d\theta^2+\sin^2\theta d\phi^2)+U(r)(d\chi+2N\cos\theta d\phi)^2,\]
with
\[U(r)=(r^2-2mr+N^2+e^2)(r^2-N^2)^{-1},\]
and this is the Wick-rotated charged Taub-NUT solution, \cite{es}, reducing to the Reissner-Nordstrom solution if $N=0$.\footnote{Since $k$ does not appear in the metric, one expects to be able to set it eual to one, and this can be achieved by adding $2\log k$ to $u$.}

The Taub-bolt solution is obtained from the Riemannian Taub-NUT solution by setting $m=5N/4$, \cite{Pa}, and that restriction applied here will give a charged version of the Taub-bolt solution.

With the Reissner-Nordstrom solution we  can obtain expressions for $V$ and $F$. First, from (\ref{w00}) and (\ref{e1}), and with the constants as identified above (and $N=0$)
\[R=\frac{2}{k}(r^2-2mr+e^2)^{1/2}\mbox{sech }x,\]
then by integrating (\ref{w2})
\[Z=-\frac{2}{k}(r-m)\tanh x,\]
(you would expect not to see the minus sign on the right in this expression; its presence is due to the clash of conventions noted above).

Next with $V_R,V_Z$ from (\ref{w00}) and the chain rule find
\[V_r=\frac{2r(r-m)}{k(r^2-2mr+e^2)}-2\frac{x}{k}\cos\theta,\]
\[V_\theta=\frac{2r}{k}\cot\theta+\frac{2}{k}(r-m)x\sin\theta,\]
which integrate to give
\be\label{e2}
V=\frac{1}{k}\left(-2(r-m)x\cos\theta+2m\log\sin\theta+2r+r_+\log(r-r_+)+r_-\log(r-r_-)\right),\ee
where
\[r_\pm=m\pm(m^2-e^2)^{1/2}.\]
To find $F$ we have
\[zF_z=W=\frac{k^2z^2}{(z^2+2az+b)},\]
whence
\be\label{e3}
F=\frac{k^2}{2(m^2-e^2)^{1/2}}\left(r_+\log(r-r_+)-r_-\log(r-r_-)\right)+f(\theta),\ee
where $f(\theta)$ is so far undetermined. To fix $f(\theta)$ recall we want $F$ harmonic in the metric $h$ of (\ref{w5}). Transforming to the $(r,\theta)$-coordinates, we find
\[h=dz^2+e^u(dx^2+dy^2)=\frac{1}{k^2}dr^2+\tD d\theta^2+\tD\sin^2\theta dy^2,\]
where $\tD=r^2-2mr+e^2$. The Laplace equation on $F$ becomes
\[k^2(\tD F_r)_r+\frac{1}{\sin\theta}(\sin\theta F_\theta)_\theta=0,\]
and imposing this on $F$ in (\ref{e3}) gives
\be\label{e4}F=\frac{k^2}{2(m^2-e^2)^{1/2}}\left(r_+\log(r-r_+)-r_-\log(r-r_-)\right)+k^4\log\sin\theta.\ee

\medskip

The key component in the  classification of 4-dimensional, Hermitian, toric, Ricci-flat metrics in \cite{BG} is the investigation of $V$ near the axis, $R=0$. They find
\[V(R,Z)=a(Z)\log R^2+ \mbox{lower order},\]
with $a(Z)$ continuous and piece-wise linear with corners at the nodes of the rod-structure (see e.g. \cite{H} for the definitions of these terms). With Reissner-Nordstrom there are two nodes, at $r=r_\pm$ or equivalently $Z=Z_\pm$ and we find
\[a(Z)=\frac{m}{k}-\frac{Z}{2},\;\;\frac{r_+}{k},\;\;\frac{m}{k}+\frac{Z}{2},\]
on the three rods, starting from negative $Z$.

If we apply the same analysis to the second harmonic function $F$ of (\ref{e4}) we find that the corresponding $a(Z)$ is discontinuous and constant on each rod.

 \item {\bf{Riemannian Kerr-Newman:}} We can take the Kerr-Newman metric from \cite{an} and replace $(t,a)$ by $(it,ia)$ to obtain it in Riemannian form
 \[g=2\tth^2(dt-a\sin^2\theta d\phi)^2 +\frac{dr^2}{2\tth^2}+\Sigma^2d\theta^2+\frac{\sin^2\theta}{\Sigma^2}(adt+(r^2-a^2)d\phi)^2\]
with
\[\Sigma^2=r^2-a^2\cos^2\theta,\;\;\tD=r^2-2mr-a^2+e^2,\tth=\frac{\sqrt{\tD}}{\Sigma\sqrt{2}}, \]
(and we're again using $\tD$ to avoid confusion with $\Delta$ in Section 3).

Comparing with (\ref{g1}) and the calculations in \cite{t2}, we may take
\[z=r-a\cos\theta,\;e^u=\tD\sin^2\theta,\;W=\Sigma^2(\tD+a^2\sin^2\theta)^{-1},\]
and then
\[x+iy=\log\left(\left(\frac{r-m-b}{r-m+b}\right)^{-a/2b}\tan(\theta/2)e^{i\phi}\right),\]
where $m\pm b$ are the roots of $\tD=0$  i.e. $b=(m^2+a^2-e^2)^{1/2}$.

 It is straightforward to verify that
  \[u_x=Z_z,\;\;e^uu_z=-Z_x,\]
  with $Z=2(r-m)\cos\theta$, so that $u$ does satisfy (\ref{w0}) but again we can't obtain $u(x,z)$ explicitly.

From (\ref{w00}) we have
  \[R^2=4e^u=4(r^2-2mr-a^2+e^2)\sin^2\theta,\]
  together with
  \[Z=2(r-m)\cos\theta,\]
from (\ref{w2}), so that $(r,\theta)$ are elliptical coordinates in the $(R,Z)$ plane. It is straightforward, if lengthy, to obtain $V(r,\theta)$: we find
  \bea\label{e5}
V&=&2(r-a\cos\theta)+2((r-m)\cos\theta-a)\log\tan(\theta/2)+2m\log\sin\theta\\\nonumber
&&+((m+b)-\frac{a}{b}(r-m)\cos\theta)\log(r-m-b)\\\nonumber
&&+((m-b)+\frac{a}{b}(r-m)\cos\theta)\log(r-m+b),\eea
which is in the same form as $V$ for Kerr in \cite{t2} but with the $b$ appropriate to Kerr-Newman.

Next we can follow the same strategy as for Reissner-Nordstrom to find $F$, obtaining
\be\label{e6}
F=\frac{1}{2b}(r_+\log(r-r_+)-r_-\log(r-r_-))+\log\sin\theta,\ee
which is very similar to (\ref{e4}).

 \medskip

 \end{itemize}

\end{document}